\documentclass[journal]{IEEEtran} %draftcls,10pt,onecolumn

\usepackage[utf8x]{inputenc}
\usepackage{cite}
\usepackage{epsfig} 
\usepackage{epstopdf}
\usepackage{graphics}
\usepackage{amsmath}
\usepackage{amssymb}
\usepackage{psfrag}
\usepackage{xspace}
\usepackage{subfigure}
\usepackage{xcolor} %Don't remove. Needs to be before footnote ... otherwise it doesn't work!
\usepackage{footnote}
\makesavenoteenv{tabular}
\makesavenoteenv{table}
\usepackage{todonotes}
\usepackage{adjustbox}
\usepackage{pdfpages}
\usepackage{tikz}
\usetikzlibrary{calc}

\begin{document}
\title{Improving practical sensitivity of energy optimized wake-up receivers: proof of concept in 65nm CMOS}
\author{\IEEEauthorblockN{Nafiseh Seyed Mazloum, Joachim Neves Rodrigues, Oskar Andersson, Anders Nejdel, and Ove Edfors}
\IEEEauthorblockN{\\Department of Electrical and Inform. Technology, Lund University, Lund, Sweden \\}
Email: \{nafiseh.seyed\_mazloum, joachim.rodrigues, oskar.andersson, anders.nejdel, ove.edfors\}@eit.lth.se}

\maketitle
\begin{abstract}
\boldmath	
We present a high performance low-power digital base-band architecture, specially designed for an energy optimized duty-cycled wake-up receiver scheme. 
Based on a careful wake-up beacon design, a structured wake-up beacon detection technique leads to an architecture that compensates for the implementation loss of a low-power wake-up receiver front-end at low energy and area costs.
Design parameters are selected by energy optimization and the architecture is easily scalable to support various network sizes.   
Fabricated in $65\mathrm{nm}$ CMOS, the digital base-band consumes $0.9\mathrm{\mu W}$ ($V_\mathrm{DD}=0.37\mathrm{V}$) in sub-threshold operation at $250\mathrm{kbps}$, with appropriate $97\%$ wake-up beacon detection and $0.04\%$ false alarm probabilities.
The circuit is fully functional at a minimum $V_\mathrm{DD}$ of $0.23\mathrm{V}$ at $f_\mathrm{max}=5\mathrm{kHz}$ and $0.018\mu$W power consumption. 
Based on these results we show that our digital base-band can be used as a companion to compensate for front-end implementation losses resulting from the limited wake-up receiver power budget at a negligible cost. 
This implies an improvement of the practical sensitivity of the wake-up receiver, compared to what is traditionally reported. 
\end{abstract}


\IEEEpeerreviewmaketitle

\section{Introduction}
\label{sec:Intro} 
\begin{figure}[t]
	\centering
	\includegraphics[width=0.38\textwidth]{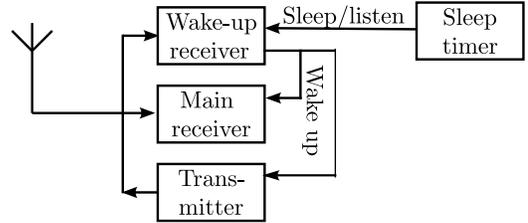}
	\caption{Simplified node block diagram. The transmitter is used both for data and wake-up beacon (WB) transmission, while the main receiver and wake-up receiver (WRx) are used for data and WB reception, respectively. The sleep timer is used when we have duty-cycled WRx scheme.}
	\label{fig:NodeBlockDiagram}
\end{figure}

Today, the success of Internet of Things has led to increasing demands on wireless sensor network (WSN) applications, through which more devices intelligently communicate with each other. 
In most of the WSN applications energy resources are severely limited both due to node sizes and the possible placement where energy resources cannot easily be replaced. 
To design a long life network it is, therefore, necessary to avoid unnecessary energy cost in the network. 
In general, in the WSNs idle channel listening is a dominant factor for energy consumption, due to their relatively low traffic intensity. 
Using an extra ultra-low power receiver, typically referred to as a wake-up receiver (WRx), dedicated for channel monitoring can significantly reduce this cost \cite{Rabaey01, Lin2004, Lont2009, Nafiseh2011}. 
There are two main approaches for how a WRx is used. 
In one, the WRx is always on, continuously listening to the channel, while in the other the WRx is duty-cycled, and only turned on periodically to listen to the channel. 
Such a WRx has limited functionality and is only used to look for potential communication, a wake-up beacon (WB), on the channel.  
When a WB is detected, the main receiver is powered up. 
A generic block diagram of an entire sensor node of this type is shown in Fig.~\ref{fig:NodeBlockDiagram}, where a sleep timer is used only if we employ duty-cycling. 
The choice of WB structure and WB detection algorithm are important in WRx schemes as they directly/indirectly influence system energy consumption.
We have proposed and analyzed detection performance of a particular WB structure in \cite{Nafiseh2012,Nafiseh2014}. 
In this paper we present the design and implementation of a WRx digital base-band (DBB) for the proposed WB. 
We show that the proposed DBB design delivers predicted performance enhancements at an energy cost low enough to make it a suitable companion to all WRx analog front-ends found in literature \cite{Pletcher2007, Pletcher2009,Durante2009,Cheng2012,Choi2012,Nilsson2013,Oh2013,Takahagi2013,Wada2013,Lee2013,Lont2011,Bae2012,Abe2014,Salazar2015,Carl2014,Huang2014,Milosiu2013,Copani2011,Marinkovic2011,Drago2010,Le2010,Hambeck2011}. 

Under realistic assumptions, a WRx typically has two orders of magnitude lower power budget than the main receiver, e.g., in the order of 10$\mu$W \cite{Rabaey01, Lin2004}. 
Since the majority of power is consumed by the WRx analog front-end, most studies focus on its design and try to minimize power consumption of this part of the circuitry \cite{Pletcher2007, Pletcher2009,Durante2009,Cheng2012,Choi2012,Nilsson2013,Oh2013,Takahagi2013,Wada2013,Lee2013,Lont2011,Bae2012,Abe2014,Salazar2015,Carl2014,Huang2014,Milosiu2013,Copani2011,Marinkovic2011,Drago2010,Le2010,Hambeck2011}. 
Simple non-coherent modulation schemes, such as on-off keying (OOK) \cite{Pletcher2007, Pletcher2009,Durante2009,Cheng2012,Choi2012,Nilsson2013,Oh2013,Wada2013,Lee2013,Salazar2015,Carl2014,Huang2014,Milosiu2013,Copani2011,Marinkovic2011,Hambeck2011}, binary frequency shift keying (BFSK)\cite{Bae2012,Abe2014,Lont2011}, pulse position modulation (PPM) \cite{Drago2010}, and pulse width modulation (PWM) \cite{Le2010,Takahagi2013}, are often used for WB transmission since they allow low-power low-complex front-end architectures. 
Extreme low-power design of such receivers, however, leads to higher noise figure and degraded sensitivity compared to a main receiver. 
The implementation/performance loss has to be compensated by increasing WB transmit energy. 
This can in principle be done by transmitting WBs with higher transmit power without extending the WB duration, or keeping the WB transmit power and making the WBs longer, e.g., by lowering data rate or applying spreading.
The first approach requires drastic increase in transmit power making it more suitable for applications where a master node without severe restricted energy source is available for WB transmission. 
In this work we are aiming for applications where all nodes have equal functionality with equal energy resources. 
To allow using energy resources as distributed as possible among nodes, we make the WBs longer by applying spreading. 
A correlator based on analogue processing \cite{Cheng2012,Choi2012} is a low-power approach commonly chosen to examine energy level of the received signal for WB detection. 
Using this approach, however, makes it difficult to distinguish between different patterns and avoid overhearing. 
Digital processing, on the other hand, allows for more flexible WB signal processing and detection algorithms. 
To prevent overhearing, identity of a node is included in the WB in the form of i) a completely unique sequence for each node \cite{Oh2013,Milosiu2013,Hambeck2011} or ii) a more structured arrangement \cite{Abe2014,Zhang2011,Nafiseh2011} consisting of, e.g., a preamble and an address part. In such an arrangement the preamble is used for synchronization purposes and identification related to individual nodes is carried in the address part. 
Correlation is performed in digital domain to detect these sequences. 
Our approach is based on the second WB structure and a corresponding digital base-band processing as it gives more flexibility to save energy by adjusting the WB to hardware characteristics and traffic requirements.
The structured WB approach also allows for shorter correlators compared to using entirely unique sequences as WB, making the signal processing more efficient both in terms of power consumption and hardware architecture flexibility.
A more detailed study of how WRx front-end characteristics influence detection performance and WB design is presented in \cite{Nafiseh2014}, where optimization is used to adjust the WB structure to minimize the energy cost of WB transmissions. 
What remains is to show, by implementation and characterization of the required base-band processing, that these schemes can deliver the performance enhancements predicted by theory without significantly increasing WRx energy consumption. 
This is what we do in this paper. 

As mentioned above, we present a DBB circuit design for a duty-cycled WRx scheme, where we show that the implementation loss of the WRx front-end, resulting from a very limited power budget, can be compensated by digital base-band processing at a negligible power consumption and area cost. 
We compare our design with those presented by others in \cite{Abe2014}, \cite{Hambeck2011} (analog/mixed signal correlator) and \cite{Zhang2011} (pure digital).
The two main differences are: i) their work assume continuous channel monitoring, while we have chosen duty-cycled operation to reduce idle listening \cite{Nafiseh2011}, and ii) our WB structure \cite{Nafiseh2012} is more flexible and allows minimization of energy cost for a wider range of node address spaces, traffic conditions, and different characteristics of the WRx analog front-end, without major changes to the DBB implementation.
This design also has the advantage of high address-space scalability, at negligible hardware cost, making the design attractive both for small and large sensor networks. 
An application specific integrated circuit (ASIC) is optimized for ultra-low voltage (ULV) operation and is characterized by measurements for different operating frequencies and a wide range of supply voltages. 
While the DBB is primarily designed and implemented for a WB with certain design parameters, chosen to compensate for the implementation loss of the WRx front-end in \cite{Carl2014}, we show that the DBB can be used as a companion to a wide range of WRx front-end design presented in literature and improve on practical sensitivities at a negligible cost in terms of power consumption. 

The paper is organized as follows. 
In Section~\ref{sec:SysDes} we give a description of the overall system operation. 
We present a hardware architecture of a WRx DBB in Section~\ref{sec:HWImpl}. 
In Section~\ref{sec:SimVal} we provide details of parameter selection for prototype implementation. Simulations are performed to evaluate receiver operating characteristics for the selected parameters. 
Measurement results from the prototype implementation are presented in Section~\ref{sec:MeasResult}. 
The performance of state-of-the-art analog front-ends is compared and discussed in Section~\ref{sec:PerfComp}. 
Conclusions and final remarks are given in Section~\ref{sec:Conclusion}. 

\section{System Description}
\label{sec:SysDes}
We design a DBB integrated circuit for a low-power duty-cycled WRx, used to search for a WB with a certain pattern, in a given time-interval. 
While low power, area efficiency and sufficient WB detection performance are essential for the DBB design itself, its integration into a larger system also has to be considered. 
In our reference system, nodes communicate according to the Duty-Cycled Wake-up receiver Medium Access Control (DCW-MAC) scheme.  
In the following we highlight some important system properties that influence our design, but for details we refer to \cite{Nafiseh2011}. 

\begin{figure}[t!]
	\centering
	\includegraphics[width=0.45\textwidth]{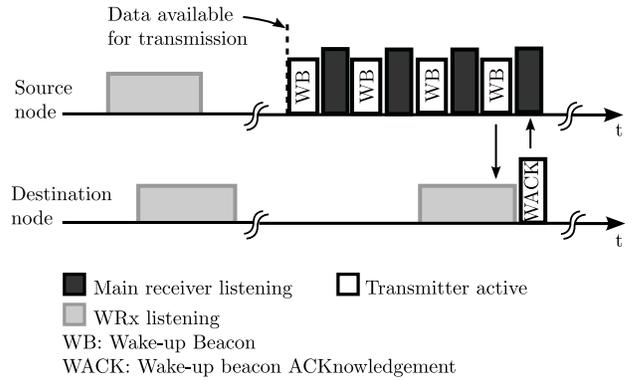}
	\caption{Simplified timing diagram of periodic wake-up beacons (WBs) and WRx duty-cycled channel listening.}
	\label{fig:DCWMAC}
\end{figure}

In the DCW-MAC, combining a low-power WRx with asynchronous duty-cycled channel listening can significantly reduce idle channel listening. 
With a low-power WRx, and the corresponding loss in performance/sensitivity, as discussed in Section \ref{sec:Intro}, the WRx needs to operate at a raw bit error rate (BER) higher than the $10^{-3}$ normally used to evaluate receiver performance. 
Since WRxs asynchronously listen to the channel, strobed WBs are transmitted, as shown in Fig.~\ref{fig:DCWMAC}, whenever data is ready for transmission. 
Using spreading and transmitting long WBs, processing gain can compensate for high BERs, improving on the practical sensitivity of the WRx.  
The WBs also carry the address of the destination node and overhearing by non-destination nodes is thereby largely avoided \cite{Estrin2002}. 
To guarantee that the WRx can hear one complete such WB, the listen interval needs to be selected long enough so that if the WRx barely misses one WB it still has the chance to capture the next one in the same listen time. Consequently, the listen interval is chosen to be at least twice the time extent of the WB, plus the time between the WBs. The time between the WBs needs to be long enough to contain a WACK packet\footnote{Whenever the WRx detects a WB, carrying its own address, the node's transmitter replies with a WB acknowledgment (WACK).}.

Ideally no errors occur during WB detection, but in a real system we have both noise and interference.
Therefore, there is a certain probability that a transmitted WB is missed, or the WRx erroneously detects a non-existing WB. 
The miss event occurs with some probability $P_\mathrm{M}^\mathrm{WB}$ and the false alarm event occurs with some probability $P_\mathrm{FA}^\mathrm{WB}$.
Both detection errors lead to unnecessary power-up of energy expensive parts of circuity, and thereby result in extra energy costs. 

All the above shows that the WRx design has an important influence on both the WB structure and the total power consumption. For details on this we refer the reader to \cite{Nafiseh2014}. In short, the use of low-power WRxs with high BER, listening to the channel asynchronously makes it important to structure the WB so that:
\begin{itemize}
	\item synchronization can be achieved,
	\item the probabilities of missing or falsely detecting a WB are kept low, and
	\item unnecessary wake-ups due to overhearing are avoided.
\end{itemize}
Here we use the WB structure from \cite{Nafiseh2012}, which fulfills the above requirements.

\subsection{Wake-up Beacon Structure}
\label{sec:WBStructure} 
The WB consist of an $M$-bit preamble and $L$-bit destination and source addresses. 
The preamble is needed to detect the presence of a WB and for time-synchronization, as the arrival time of the WB in the WRx listen interval is unknown. 
For simplicity, the preamble is selected to be identical for all nodes since uniqueness is provided by the address part. 
The destination address is used to avoid activating non-destination nodes, while the source address is used in the destination address field of the WACK. 

For accurate time-synchronization, the preamble should have good autocorrelation properties and it should be long enough to compensate for the high BER of the front-end.
For the destination and source address fields, we do not need the autocorrelation properties, but the high BER still has to be compensated. We do this by $K$-bit spreading of each address bit, using an arbitrary code, resulting in a total of $2KL$ bits for both addresses. This leads to a $M+2KL$ bit WB where energy optimization can be done over $M$ and $K$. The optimal $M$ and $K$ depend on system parameters like traffic conditions, delay requirements and network size. It is therefore of interest to make a DBB implementation, as done below, that can be easily adjusted to different $M$, $K$ and $L$. Typical ranges, when energy optimizing networks with up to $L=16$ address bits and front-end BERs as high as 0.15, are $M \lesssim 60$ and $K \lesssim 10$, with $M$ roughly ten times larger than $K$ for individual optima \cite{Nafiseh2014}. 

\section{Digital Base-band Hardware Architecture}
\label{sec:HWImpl}
\begin{figure}[t!]
	\centering
	\includegraphics[width=0.42\textwidth]{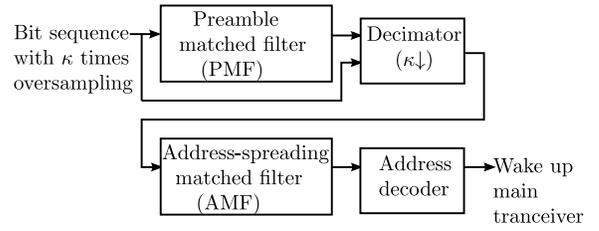}
	\caption{Digital base-band (DBB) block diagram.}
	\label{fig:WRx}
\end{figure}
After establishing our WB structure, we propose a hardware architecture for the DBB processing. There is, however, a point that we deliberately avoided in the above discussions that need to be addressed. We implicitly assumed that there was a bit-synchronization between transmitter and receiver, which of course is not the case. We handle this by assuming an oversampling factor, where the analog front-end of the WRx delivers bit-decisions at $\kappa$ times the actual bit rate. The DBB therefore has to perform its processing at $\kappa$ times the channel bit rate. Given this, the task of the DBB is to search for the presence of a WB in this bit sequence.

We have chosen the block diagram shown in Fig.~\ref{fig:WRx}, consisting of a preamble matched filter (PMF), a decimator, an address-spreading matched filter (AMF), and an address decoder.
All MFs in our design are finite impulse response (FIR) filters with the transfer function $F(z)=\sum_{i=0}^{J-1}f_\mathit{i}\,z^\mathit{-i}$ where $J$ is the number of filter taps, and the values of the filter impulse response $f_\mathit{i}$s are the reversed known sequence, i.e., the preamble, the address spreading, and the node address, we are looking for.
All these MFs are also followed by a comparator acting as decision device and, for simplicity, we include this component when using the term MF. 
Feeding the input $x[n]$ in the form of a bit stream to the MF, the filter output $y[n]$ becomes 
\begin{equation}
y[n]=\sum_{k=0}^{J-1}f_\mathit{k} \, x[n-k],
\label{eq:MF}
\end{equation}
and when $y[n]$ is larger than a predefined threshold $\gamma$, we assume a detection.
The DBB searches the received bit sequence for the WB, based on above principle, in the following steps.
First the PMF is used to search for the preamble, at a $\kappa$ times oversampling, since it is the part of the WB designed for synchronization.
Whenever the output of the PMF exceeds a certain threshold a preamble is detected. 
The maximum peak, indicating the correct clock phase, is found among $\kappa$ successive samples for improved time synchronization.  
After preamble detection, the input sequence is passed to a decimator ($\kappa$$\downarrow$), and the rest of the processing is performed at channel bit rate.
The remainder of the bit-sequence is fed to the address-spreading matched filter (AMF), where the individual address bits are detected by correlating the sequence with the address spreading sequence. 
At this stage the DBB knows the synchronization and performs correlation only once per address bit.  
Finally, the detected address bits are collected by the address decoder and compared against the node address.  
If there is a match, the main transceiver is powered up. 

With the proposed architecture, the PMF and the AMF are identical in all nodes of a network. 
Only the address decoder needs to be programmed with the respective node addresses. 
The advantages of our WB structure, and DBB design, over the structures proposed in \cite{Hambeck2011} and \cite{Zhang2011} are that the selection of WB pattern and address code is not limited to a certain code-book, and the programmable address decoder enables a large address-space scalability.
For instance, to scale a network size from $256$ to $1024$ nodes, we only need to increase the address decoder length (from $8$ to $10$ bits), while the PMF and the AMF can remain unchanged. 
Moreover, accurate time-synchronization provided by oversampling and using preambles with sharp peaks allows us to process the address part of the WB without oversampling, leading to shorter correlators for address detection. 
Furthermore, the DBB design is improved, over a previous design \cite{Nafiseh2011}, by detecting the address bits using the AMF and the address decoder, instead of using one MF for the entire address field.
The new design, realized in hardware as binary-input MFs, leads to both a shorter critical path and smaller area and, consequently, less leakage energy. 
Latency is the same for both structures since the number of clock-cycles before the DBB decides if a WB is present remains the same.  

What now remains is to specify in more detail the implementation of the MFs and the decimator.
\subsection{Matched filters} 
\label{sec:MF}
\begin{figure}[t]
	\centering
	\includegraphics[width=0.48\textwidth]{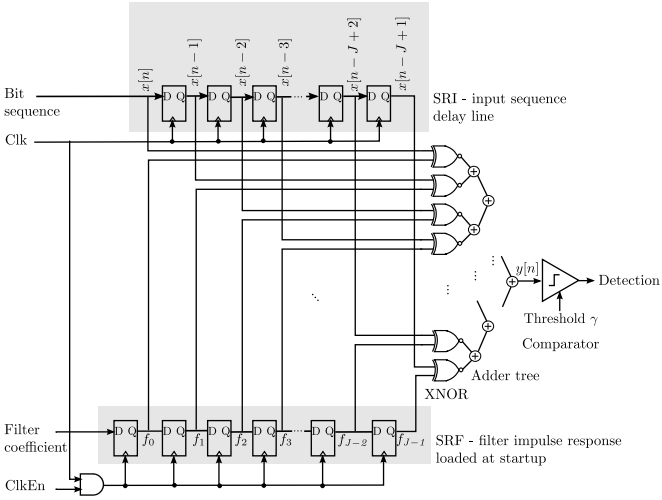}
	\caption{Hardware mapping of a generic binary-input matched filter, consisting of two shift registers (SRI and SRF), XNORs, an adder tree, and a comparator.}
	\label{fig:HWMappingWBStr}
\end{figure}
We describe a generic hardware mapping of a binary-input MF used to compute \eqref{eq:MF}, since MFs are the main building blocks of our DBB design. 
The differences in the deployment of the MF for the PMF, AMF, and address decoder are further explained. 

As depicted in Fig.~\ref{fig:HWMappingWBStr} the binary-input MF is implemented using two shift registers, one for storing the filter impulse response (SRF) and one acting as a delay line for the incoming bits (SRI).
% and a programmable filter sequence.
During the initialization phase, the clock enable ClkEn is set to one and the reversed known sequence, i.e., the preamble, the address spreading or the node address, is fed to the SRF. All values of the filter impulse response, $f_0 ... f_\mathit{J-1}$, are stored in the SRF after $J$ clock cycles. 
All bits in the sequence $x[n] ... x[n-J+1]$ necessary to calculate the output are available by feeding the incoming bits to the SRD, which is shifted one bit at each clock cycle.
Correlation of the input signals ($x[n] ... x[n-J+1]$) with the values of the filter impulse response ($f_0 ... f_\mathit{J-1}$) is performed at bit-level, where XNORs are the first stage to create filter tap outputs. 
Summation of these outputs is then realized by a fully balanced adder tree. 
Thus, \textit{idle time} of the gates is kept low and, consequently, energy dissipation due to leakage reduces. 
The adder tree is composed of half-adders (mirror architecture), taken form the standard-cell library.

Using the above MF hardware mapping, the differences between PMF, AMF and address decoder are in clock rate, length of SRs, number of filter taps, and comparator threshold level. 
Both the number of SRs and filter taps for the PMF are $\kappa$ times the preamble length $M$, since the PMF receives the oversampled bit sequence. 
The PMF is clocked at $\kappa$ times the bit rate.
The AMF and the address decoder are placed after the decimator and receive the bit sequence at a normal channel bit rate. 
This means that the number of SRs and filter taps for the AMF and the address decoder are equivalent to the length of the address-spreading $K$ and the number of address bits $L$, respectively. 
The decision level in the PMF can vary in the range $\left[ 0~~(4M-1)\right] $ and depends on performance requirements and front-end BER.
In the AMF, responsible for address-bit detection, we set the threshold level to the midpoint $\lceil K/2 \rceil$. Since we require all address bits to match the node address, the address decoder threshold is set to $L$.

\subsubsection*{Sub-$V_\mathrm{T}$ characterization of the MFs}
Sub-$V_\mathrm{T}$ characterization of a single MF in \cite{Nafiseh2012} shows that maximum operational frequency varies only slightly with filter length. This agrees with the fact that the critical path primarily depends on the depth of balanced adder tree, growing only logarithmically with filter length. Energy per clock cycle and area, on the other hand, are highly dependent on the filter length and scale roughly linearly. Given the experience discussed in Section~\ref{sec:SysDes}, with $M \approx 10 K$, DBB characteristics will be dominated by the large PMF of length $\kappa M$.

\subsection{Decimator}
\begin{figure}[t]
	\centering
	\includegraphics[width=0.48\textwidth]{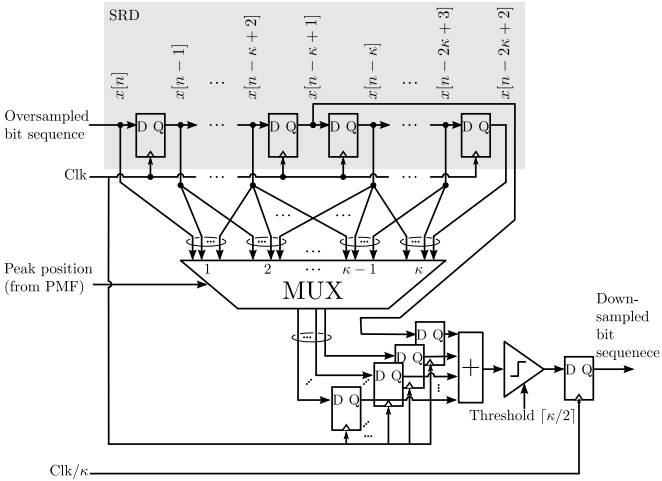}
	\caption{Hardware-mapping of a decimator consisting of a shift register (SRD), a multiplexer, an adder, and a comparator.}
	\label{fig:Decimat}
\end{figure}

After preamble detection and bit-synchronization, we do not need to continue at the oversampled rate and can operate at normal bit rate when de-spreading and detecting node address. This, as previously mentioned, saves energy and reduce area compared to operating directly on the oversampled sequence. 
The decimator in Fig.~\ref{fig:WRx} is, therefore, used to perform the down-sampling. 
Using the position of bit-timing/clock-phase from the PMF output, the decimator down-samples the sequence by adding $\kappa$ oversampled bits at a time. 
The result of this is thresholded to decide whether the down-sampled bit is a zero or one.   
Figure \ref{fig:Decimat} shows the hardware implementation of a decimator, consisting of a ($2\kappa-1$)-bit SR, indicated by SRD, a ($\kappa-1$)-bit $\kappa$ to $1$ multiplexer, an adder and a comparator. 
With $\kappa$ times oversampling, there are $\kappa$ possible correct bit-timing/clock-phase for the decimator to perform the summation of the incoming oversampled bits. 
To have access to all $x[n] ... x[n-(2\kappa+2)]$ samples needed to calculate the output for any of the clock phases, the ($2\kappa-1$)-bit SRD is used.
The input $x[n]$ is directly connected to the oversampled bit sequence and the SRD stores all above samples by shifting the incoming bits every clock cycle.  
The multiplexer inputs are then fed with $\kappa$ choices of incoming sample sequences, grouped based on the possible clock phase. 
The peak position of the PMF output is fed to the multiplexer control input. 
The control output in return feeds through the correct ($\kappa-1$) samples to the adder for further processing.
Since input sample $x[n-\kappa+1]$ is present in all sums, independent of the peak position, it is fed directly to the adder instead of through the multiplexer. 
This allows us to use a multiplexer with a smaller size, saving both on energy and area. 
The comparator output is set to zero if the output of the adder is smaller than $\kappa/2$ while it is set to one for the other values. 

Due to its small size, contributions from the decimator on total DBB sub-$V_\mathrm{T}$ characteristics will be negligible for reasonable parameter choices.
\section{Parameter Selection}
\label{sec:SimVal} 
Both WB detection performance and power consumption of the DBB implementation are of importance to the overall evaluation of the proposed architecture.
In this section we select implementation parameters to obtain sufficient detection performance, while power consumption is discussed in the next section.

The WB detection performance has been extensively studied from theoretical point of view in \cite{Nafiseh2014}, where WB parameters are optimized for different front-end characteristics and network sizes. Ranges of resulting parameters were discussed briefly in Section~\ref{sec:SysDes}.
As a proof of concept we implement the DBB with realistic design parameters from the \emph{Ultra-Portable Devices} project at the Department of Electrical and Information Technology, Lund University \cite{Henrik2012},\cite{Henrik2014} and a particular low-power analog front-end \cite{Carl2014} in mind. The analog front-end is designed for operation at $2.4\mathrm{GHz}$ and $250\mathrm{kbps}$ on-off keying carrying Manchester coded bits. 
Using a passive mixer together with a ring oscillator, the analog front-end down-converts the received RF signal to IF.
The envelope of the IF signal is detected and filtered by a band-pass filter that reduces noise and interference outside the expected range, including DC from constant envelope signals. 
Using a simple non-coherent signal energy detector, channel bits are detected at $\kappa=4$ times oversampling and fed to the DBB.
The combination of passive mixer with three-phase mixing and complementary IF amplifiers improves efficiency
resulting in $-88\mathrm{dBm}$ sensitivity at $10^{-3}$ BER and $50~\mu$W power consumption. 
For more details and a block diagram of this particular analog front-end see \cite{Carl2014}. 
We consider a network with maximum $256$ nodes ($L=8$) and a channel BER of $0.15$. The high BER can be traced back to operating the analog front-end at a practical sensitivity level equal to that of the main receiver, $-94\mathrm{dBm}$. 
This corresponds to a need to improve the practical sensitivity by $6\mathrm{dB}$.
Along the lines described in Section~\ref{sec:WBStructure}, energy optimized WB parameters fall in the range of $M=31$ and $K=7$, for this scenario. The particular value $M=31$ is related to lengths of $m$-sequences with good autocorrelation properties \cite{Martin1977}. Applying the factor-four oversampling, Manchester coding of bits, and rounding up to the nearest power of two, gives PMF and AMF lengths of $256$ and $16$ bits, respectively.

\begin{figure}[t]
	\centering
	\includegraphics[width=0.45\textwidth]{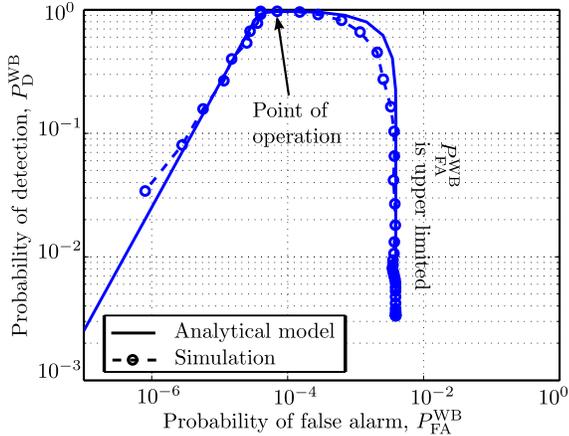}
	\caption{Simulated and calculated receiver operating characteristics, ROCs, for a Manchester-coded wake-up beacon (WB) with a preamble of length $M=62$, $L=8$ bit addresses and address spreading $K=14$. $P_\mathrm{FA}^\mathrm{WB}$ is calculated assuming that an interfering WB is always present during channel listening.}
	\label{fig:DBBROC}
\end{figure}

While thresholds for the AMF and address decoder are fixed, the DBB performance, in terms of detection $P_\mathit{D}^\mathrm{WB}=1-P_\mathit{M}^\mathrm{WB}$ and false alarm $P_\mathrm{FA}^\mathrm{WB}$ probabilities, changes with the PMF threshold level. Using the analytical framework from \cite{Nafiseh2014} with parameters as specified above, we show the receiver operational characteristics (ROC) of the DBB in Fig.~\ref{fig:DBBROC}. The analytical curve shows the ROC for ideal correlation properties, while simulations are performed for the non-ideal Manchester-coded 31-bit m-sequence used in the implementation. As can be seen, the simplified analysis and realistic simulations agree well. The chosen point of operation for our implementation is a PMF threshold at $92\%$ of the maximum filter output, which provides $97\%$ WB detection probability and a low WB false alarm probability, in the order of $10^{-4}$. Both probabilities are given per listen interval, which is set to the minimal value of twice the WB length.

\section{Measurement Results}
\label{sec:MeasResult}
\begin{figure}[t]
	\centering
	\includegraphics[width=0.42\textwidth]{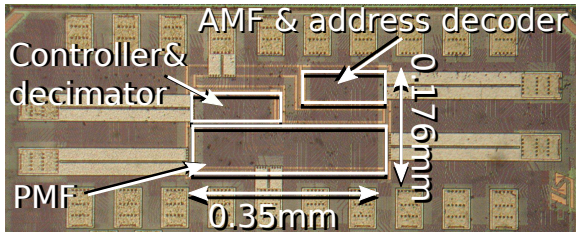}
	\caption{Chip micro-photograph.}
	\label{fig:DiePhoto}
\end{figure}

The DBB is fabricated in a $65\mathrm{nm}$ CMOS technology.    
Figure \ref{fig:DiePhoto} shows the chip micro-photograph. 
The area of the integrated circuit, including peripheral access, is $0.062\mathrm{mm^2}$.
The functionality of the fabricated chip has been verified by connecting the output of an analog front-end \cite{Carl2014}, from the \emph{Ultra-Portable Devices} project, to the DBB input. 

\begin{figure}[t]
	\centering
	\includegraphics[width=0.4\textwidth]{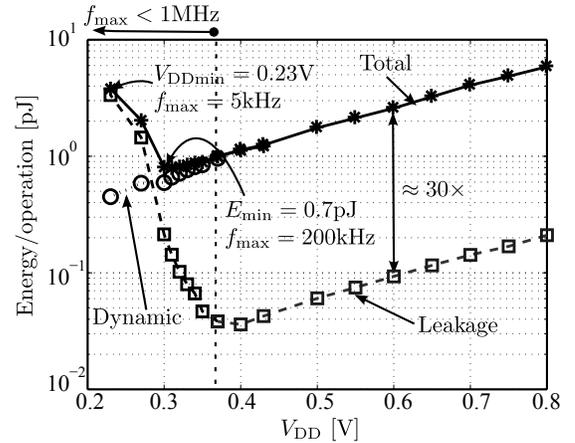}
	\caption{Measured energy vs. $\mathrm{V}_\mathrm{DD}$ at room temperature. The dashed vertical line indicates the lowest  $\mathrm{V}_\mathrm{DD}$ at which the target 250kbps can be sustained.}
	\label{fig:MeasDBB}
\end{figure}

Fig.~\ref{fig:MeasDBB} shows the result of our measurements. The dashed vertical line at $V_\mathrm{DD}=0.37\mathrm{V}$ indicates the lowest supply voltage at which the $250\mathrm{kbps}$ ($f=1\mathrm{MHz}$ with $\kappa=4$ oversampling) can be maintained. At $250\mathrm{kbps}$ operation, we see that leakage is negligible at $30\times$ below dynamic energy dissipation, even if the circuit is not operated at maximum operating frequency ($f_\textrm{max}$). Dissipating $0.9\mathrm{pJ}$/operation at $1\mathrm{MHz}$ gives a power consumption of $0.9\mathrm{\mu W}$. This shows that the presented DBB design compensates for the implementation loss of the low-power analog front-end ($50\mathrm{\mu W}$) \cite{Carl2014} at negligible power consumption. 

\begin{figure}[t]
	\centering
	\includegraphics[width=0.37\textwidth]{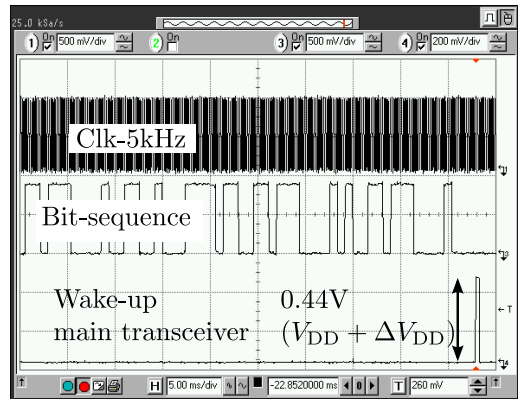}
	\caption{Oscilloscope measurement at min. $\mathrm{V}_\mathrm{DD}$ of $0.23$V@$5$kHz at room temperature. A $\Delta \mathrm{V}_\mathrm{DD}$ higher supply voltage is needed to drive the pads. 
	}
	\label{fig:MeasDBBOsc}
\end{figure}

Measurements in the sub-$V_\mathrm{T}$ region, at $f_\textrm{max}$, show an energy minimum of $E_\mathrm{min}=0.7\mathrm{pJ}$/operation at $\mathrm{V}_\text{DD}=0.31\mathrm{V}$ ($f_\mathrm{max}=200\mathrm{kHz}$), giving a power consumption of $140\mathrm{nW}$. The DBB is fully functional down to lowest supply voltage $V_\mathrm{DDmin}=0.23\mathrm{V}$ ($f_\mathrm{max}=5\mathrm{kHz}$) which, to the authors best knowledge, is lower than any number published in literature. 
While Fig.~\ref{fig:MeasDBB} shows measurement results at room temperature, measurements at body temperature show that minimal energy per operation, $V_\mathrm{DD}$ at minimal energy, and lowest operational $V_\mathrm{DD}$, all increase by less than 20\%. This shows that despite the 20\% increase, the DBB power consumption is still negligible compared to the analog front-end power consumption.
Fig.~\ref{fig:MeasDBBOsc} displays the oscilloscope measurements of the circuit at V$_\mathrm{DDmin}$ at room temperature.
The power consumption at this point is $18\mathrm{nW}$. 

Comparison with previous DBB designs\footnote{Among the solutions found in literature \cite{Cheng2012,Choi2012,Abe2014,Hambeck2011,Milosiu2013,Oh2013,Zhang2011} we have chosen to directly compare to those where WB detection performance has been reported, namely \cite{Abe2014}, \cite{Hambeck2011} and \cite{Zhang2011}.} is shown in Table~\ref{tab:LitWRxPerDBB}. 
Our WB structure is more flexible than previous designs, allowing arbitrary WB pattern and address spreading.   
The selected WB structure and DBB processing results in two to three orders of magnitude lower $P_\mathrm{FA}^\mathrm{WB}$ than in  \cite{Abe2014}, \cite{Hambeck2011} and \cite{Zhang2011}, while $P_\mathrm{D}^\mathrm{WB}$ remains on the same level.
Moreover, this work outperforms \cite{Zhang2011} both in terms of power consumption and lowest supply voltage at which it is fully functional.  
The power consumption of this work and the efficient analog hybrid solution \cite{Hambeck2011}, not characterized for low supply voltage, are comparable at their normal $V_\mathrm{DD}$s when taking the difference in data rates into consideration. 
By using the WRx for address detection, we also avoid the energy consuming process of waking up the power-hungry main receiver to check the address \cite{Abe2014} of each WB.
Moreover, our DBB is optimized for duty-cycled WRxs and by optimizing sleep time of our WRx the average power consumption can go down drastically and ideally approach the WRx sleep power, which for our design is $0.5\mathrm{nW}$. 
Previous studies on the DBB are optimized for always-on processing and do not use a sleep mode to reduce average power consumption.
\begin{savenotes}
\begin{table*}[!t]
	\small
	\vspace*{-3mm}
	\centering
	\caption{Comparison to previous work.}
	\begin{adjustbox}{width=0.6\textwidth}
		\vspace*{-3mm}
		\renewcommand\arraystretch{1.3} 
		\begin{tabular}{l|c|c|c|c}
			                                                             \multicolumn{5}{c}{}                                                               \\ \hline
			\textbf{Parameter}                                      & \cite{Abe2014} \footnote{When Energy Detection Packet (EDP) is detected by a low-power receiver, a more power hungry receiver is powered up to detect Address Detection Packet
				(ADP). Presented power consumption is only for low-power wake-up receiver.} & \cite{Hambeck2011} & \cite{Zhang2011} &   This work    \\ \hline\hline
			WB type                                                 &            EDP +             &       Unique       & Preamble+Sync.+  &   Preamble +   \\
			                                                        &             ADP              &      sequence      &  Codebook addr.  & Spread address \\ \hline
			$P_\mathit{D}^\mathrm{WB}$, $P_\mathit{FA}^\mathrm{WB}$ &         0.999, 1E-3          &     0.99, 1E-3     &   0.98, 2.8E-2   &   0.97, 4E-5   \\
			(per wake-up beacon)                                    &          Always-ON           &     Always-ON      &    Always-ON     &  Duty-cycled   \\ \hline
			Power cons. [$\mu$W] and                                &             44.2             &        0.4         &       3.72       &      0.9       \\
			Data rate [kbps]                                        &              50              &        100         &       200        &      250       \\
			@ V$_\mathrm{DD}$ [V]                                   &             0.7              &        1.0         &       1.2        &      0.37      \\ \hline
			Power cons. [$\mu$W]                                    &              NA              &         NA         &       0.9        &     0.018      \\
			@ V$_\mathrm{DDmin}$ [V]                                &                              &                    &       0.6        &      0.23      \\ \hline
			Technology [nm]                                         &              65              &        130         &        90        &       65       \\ \hline
			Area [mm$^2$]                                           &          $\sim$0.42          &     $\sim$0.12     &       0.1        &     0.062      \\ \hline
		\end{tabular}
	\end{adjustbox}
	\label{tab:LitWRxPerDBB}
\end{table*}
\end{savenotes}

\section{Discussion}
\label{sec:PerfComp}
\begin{figure}[t]
	\centering  
	\begin{tikzpicture}
	\node[anchor=south west,inner sep=0] at (0,0) {\includegraphics[width=0.45\textwidth]{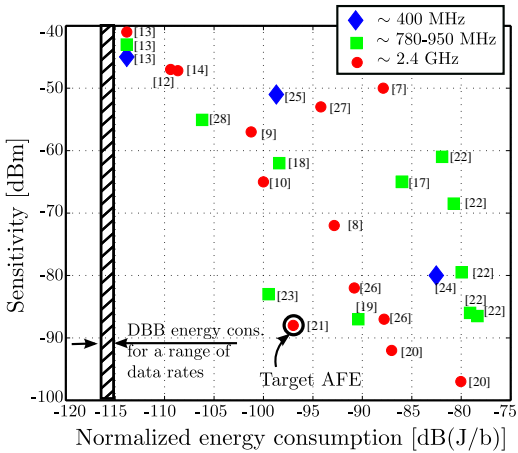}};
	\draw (2.5,5.608) node {\tiny \cite{Oh2013}};
	\draw (2.5,5.43) node {\tiny \cite{Oh2013}};
	%\draw (3,5.43) node {\tiny \cite{Choi2012},~};
	\draw (2.5,5.25) node {\tiny \cite{Oh2013}};
	\draw (3.2,5.1) node {\tiny \cite{Takahagi2013}};
	\draw (2.75,4.95) node {\tiny \cite{Nilsson2013}};
	\draw (4.5,4.75) node {\tiny \cite{Marinkovic2011}};
	\draw (5.88,4.85) node {\tiny \cite{Pletcher2007}};
	\draw (5.1,4.6) node {\tiny \cite{Le2010}};
	\draw (4.15,4.25) node {\tiny \cite{Durante2009}};
	\draw (6.7,3.95) node {\tiny \cite{Huang2014}};
	\draw (4.55,3.87) node {\tiny \cite{Bae2012}};
	\draw (4.3,3.6) node {\tiny \cite{Cheng2012}};
	\draw (6.15,3.6) node {\tiny \cite{Lont2011}};
	\draw (6.9,3.33) node {\tiny \cite{Huang2014}};
	\draw (5.3,3.05) node {\tiny \cite{Pletcher2009}};
	\draw (3.5,4.45) node {\tiny \cite{Hambeck2011}};
	\draw (6.5,2.2) node {\tiny \cite{Copani2011}};
	\draw (5.5,2.2) node {\tiny \cite{Drago2010}};
	\draw (4.4,2.1) node {\tiny \cite{Milosiu2013}};
	\draw (4.8,1.7) node {\tiny \cite{Carl2014}};
	\draw (6.05,1.35) node {\tiny \cite{Salazar2015}};
	\draw (6.95,0.95) node {\tiny \cite{Salazar2015}};
	\draw (5.45,1.95) node {\tiny \cite{Abe2014}};
	\draw (5.9,1.8) node {\tiny \cite{Drago2010}};
	\draw (7,2.4) node {\tiny \cite{Huang2014}};
	\draw (6.9,2.05) node {\tiny \cite{Huang2014}};
	\draw (7.15,1.9) node {\tiny \cite{Huang2014}};
	\end{tikzpicture}
	\caption{Performance comparison of WRx analog front-ends found in literature, in terms of energy consumption and sensitivity. The measured range of energy consumption of the proposed DBB, for 10 to 250 kbps, is shown as a hatched region.}
	\label{fig:WRxComp}
\end{figure}
We have shown, in the previous section, that our proposed DBB outperforms the existing DBB solutions in terms of power consumption and detection performance. 
We have also shown that the flexibility of the proposed DBB allows adjusting the design without significant changes in hardware architecture or power consumption. 
In this section we show that, without significantly increasing power consumption of the WRx, the proposed DBB can be connected to a wide range of optimized WRx analog front-ends and improve on their performance.  

Low-power analog front-end WRx design, as mentioned in Section \ref{sec:Intro}, has been a popular and active research area for more than a decade \cite{Pletcher2007, Pletcher2009,Durante2009,Cheng2012,Choi2012,Nilsson2013,Oh2013,Takahagi2013,Wada2013,Lee2013,Lont2011,Bae2012,Abe2014,Salazar2015,Carl2014,Huang2014,Milosiu2013,Copani2011,Marinkovic2011,Drago2010,Le2010,Hambeck2011}. 
Depending on the target applications and parameter choices, these designs are optimized to operate at different data rates and operating frequencies making a trade-off between sensitivity, power consumption and resulting wake-up delay. 
The performance, sensitivity vs. energy consumption, of existing analog front-ends (including the targeted AFE from the \emph{Ultra-Portable Devices} project) is presented in Fig.~\ref{fig:WRxComp}. To be able to compare these designs reasonably fair we normalize power consumptions to their corresponding data rates.
The hatched region to the left is the range of energy consumption measured for our DBB, for data rates between 10 and 250 kbps. 
This covers most data rates at which the analog front-ends are operable.
The DBB in itself does not have an associated sensitivity and the region therefore extends across all sensitivity levels.
We can see that our DBB will essentially not increase the total WRx power consumption, since its energy consumption is significantly lower, often orders of magnitude, than that of the analog front-ends.  
This shows that our proposed DBB can compensate for implementation losses and improve the practical sensitivity of the target analog front-end, for which it was designed, at a negligible energy cost and it can do the same for a wide range of analog front-ends found the literature. 
The improvement in terms of practical sensitivity for our target analog front-end, as shown in Section \ref{sec:SimVal}, is $6\mathrm{dB}$ and for the same requirement on detection performance the same improvement can be achieved for the other analog front-ends as well.

\section{Conclusions}
\label{sec:Conclusion} 
A digital base-band design for a duty-cycled WRx in $65\mathrm{nm}$ CMOS is presented. 
With adequate level of detection performance, the total power consumption of the digital base-band ($0.9 \mathrm{\mu W}$) is negligible in comparison with our analog front-end power consumption ($50\mathrm{\mu W}$) \cite{Carl2014}. 
This shows that implementation loss resulting from aggressive power savings in the
analog front-end can be efficiently compensated with digital base-band processing. 

%\bibliographystyle{IEEEtran}
%\bibliography{refs}


\end{document}